# The Relationship Between Environmental Regulation and Urbanization: a panel data analysis of Chinese prefecture-level cities

Chao Zhang[1], Yulin Lu[2]
[1]Kennedy School, Harvard University
[2]School of Economics, Fudan University

## 1. Introduction

Since the Industrial Revolution, the world economy has experienced rapid development, and China's economy has also achieved an unprecedented takeoff in the past. Behind the economic growth, population surge, and continuous improvement of people's living standards lies the enormous consumption of fossil energy and environmental pollution. This kind of pollution has caused irreparable damage to the world. The most concerned environmental issue globally at present is the global warming caused by carbon dioxide emissions. China is in a stage of rapid development, and as the largest developing country, China's development path has a significant impact on global climate change. At the same time, the global community also puts pressure on China to limit carbon dioxide emissions. To address energy shortages and environmental issues, countries around the world have introduced corresponding energy and environmental regulations. Due to different culture and government systems, the effects of energy and environmental regulations in various countries are also different. Therefore, it is still necessary to discuss China's energy and environmental regulations. On the one hand, environmental regulations implemented to save energy and reduce emissions will have a direct impact on the environment, and on the other hand, they will also affect the flow of enterprises and population, thereby indirectly affecting urbanization, which may change the original intention of energy and environmental regulations. In the past, there were few literature that investigated the relationship between energy and environmental regulations and urbanization. Most of the literature focused on the relationship between energy and environmental regulations and enterprise migration, and employment. There is no literature that directly discusses the relationship between energy and environmental regulations and the urbanization rate. China's urban development, due to the existence of the Hukou system, is still somewhat different from the world's urban development path. Therefore, this paper hopes to use the environmental policy in the 11th Five-Year Plan, combined with Chinese Hukou system, to analyze the relationship between energy and environmental regulations and the flow of urban population.

This paper is structured as follows: Section 2 is the Literature Review; Section 3 analyzes the theoretical crowding-in and crowding-out effects of environmental regulation on urban population mobility; Section 4 introduces the econometric model and data sources; Section 5 presents the basic regression results; Section 6 discusses relevant mechanisms; Section 7 concludes with policy implications.

## 2. Literature Review

The relationship between environmental regulation and corporate location as well as employment has been extensively studied by scholars (Berman & Bui, 2001; Hazilla & Kopp, 1990; Hollenbeck, 1979; Wendling & Bezdek, 1989); Wendling and Bezdek (1989), among which the Pollution Haven Hypothesis proposed by Copeland and Copeland and Taylor (1994) is quite famous. This hypothesis predicts that international free trade will lead to the transfer of pollution-intensive industries from developed countries with strict environmental regulations to developing countries with more lenient regulations. Subsequently, a large number of scholars have tested this hypothesis using data from various countries. Dean et al. (2009) found in China that regions with looser environmental regulations have a higher proportion of pollution-intensive industries, but there is no evidence that these pollution-intensive enterprises come from developed countries. Kahn and Mansur (2013), using data from the U.S. manufacturing industry, found that pollution-intensive enterprises tend to be established in areas with more relaxed air pollution regulations. They indicated that U.S. energy-intensive industries prefer to be located in areas where electricity prices are cheaper, while pollution-intensive industries prefer areas with looser environmental regulations. Cole et al. (2010), using data from Japan, studied the Pollution Haven Hypothesis and found that the phenomenon is more severe when enterprises have trade with developing countries, have higher pollution discharge costs, and when factories are not easy to relocate. The Chinese government initiated the "Double Reduction" policy in 1998, and Cai et al. (2016), using a difference-in-differences model, found that the policy reduced foreign investment. At the same time, foreign investment is not sensitive to environmental regulations in countries with better environmental protection than China, but is sensitive to regulations in countries with worse environmental conditions than China. Chen and Cheng (2017) also studied the behavior of enterprises under the "Double Reduction" policy and found that the targeted areas have seen significant environmental improvements, mainly due to the closure of pollution-intensive enterprises within the region, but this has also led to an increase in new pollution enterprises in non-targeted areas. In addition, the change in the evaluation criteria for local governments in China at the beginning of the 21st century also facilitated the implementation of this policy. Shen et al. (2017) used a spatial autoregressive model to study the impact of environmental regulation on the total amount of pollutants discharged by enterprises and the total number of environmental violators, and found that the intensity of local environmental regulation significantly reduced the discharge of pollutants by enterprises. However, the greater the intensity of environmental regulation in neighboring areas, the more detrimental it is to the local environment. The reason is that enterprises in neighboring areas, due to strict environmental policies, will relocate to areas with relatively less strict local environmental regulations, thereby increasing the number of polluting enterprises in the local area and causing the deterioration of the local environment.

Another group of scholars has analyzed the relationship between environmental regulation and employment. Curtis (2017) used a triple difference method to study the regulation of nitrogen dioxide and found that the regions implementing this regulation lost 82,000 jobs. Higher electricity price possibly was the underlying mechanism of this regulation's impact. Liu et al. (2017) used firm-level data to analyze the relationship between water pollution regulation and employment in the textile industry, and the results showed that firms subject to stricter new regulations reduced their labor demand by about 7%. Foreign and state-owned enterprises were not significantly affected, while local-owned enterprises were significantly affected. Yip (2018) used individual-level data to study the impact of Canada's carbon tax on employment and found that the policy increased the unemployment rate of moderately and less-educated males by 1.4% and 2.4%, respectively. Workers affected by the policy either had to look for short-term part-time work, or some left the labor market.

The literature discussed primarily concerns the relationship between environmental regulation and corporate location and employment. Residents also choose their cities of residence based on personal preferences, a process known as "voting with their feet." On one hand, residents migrate from rural areas to urban areas; on the other hand, urban residents also make choices about where they live within cities. Tiebout (1956) pointed out that residents would weigh public facilities, living environments, and personal preferences, ultimately choosing their place of residence through the mechanism of "voting with their feet". Indeed, as mentioned earlier, industrialization has played a significant role in the development of urban areas. The job opportunities created by urban industrialization are also a major reason for the migration of people from rural to urban areas. Wong et al. (2006) found that the rapid economic development of China's prosperous cities has led to a massive migration of surplus labor from inland areas to coastal cities for development. In the past, job opportunities dominated residents' choices of cities. However, as society develops and people's environmental awareness increases, research on the harms caused by environmental pollution has increased. Shinsuke and Tanaka (2015) analyzed the "Dual Control Zone" policy and infant mortality and found that the infant mortality rate in the control area decreased by 20% during this period. The reduction in infant mortality was mainly concentrated in the neonatal period, and he believed there might be an underlying pathological mechanism. Many other scholars have also found that environmental pollution can cause physiological harm to people, especially infants and young children (Chen et al., 2013; Ebenstein et al., 2017; Greenstone & Hanna, 2014). Therefore, an increasing number of residents also consider environmental quality as an important factor in choosing where to live.

When residents are initially affected by environmental pollution, they can adopt pollution avoidance measures, such as purchasing protective masks, using air purifiers, and water purifiers to reduce the impact of environmental pollution on themselves (Ito & Zhang, 2019; Zhang & Mu, 2017). However, when environmental pollution becomes more severe, more and more residents will consider the option of immigration. Qin and Zhu (2018) conducted a study on China's air pollution and residents' willingness to immigrate, finding that when the daily air pollution index is higher, the frequency of residents' searches for immigration will significantly increase the next day. The impact of environmental regulation on population mobility is not uniform. In fact, due to the different ideas and economic strengths of the labor force, environmental pollution has a significant heterogeneous impact on different types of labor (Scully, 1996). Glaeser et al. (2000) found that as the mobility of enterprises increases, the development of cities increasingly depends on the benefits they can bring to consumers for living. He found that cities with better infrastructure and environment develop faster. Moreover, the cost of living in cities rises more than wages, which means living in a good city can bring additional benefits to residents. Zheng et al. (2009) pointed out that with the improvement of the mobility of China's labor force, the mechanism of "voting with their feet" to choose a city of residence will be more and more adopted by Chinese workers, and urban environment, education, and other factors will be given higher weight. The impact of environmental regulation on different labor forces is different. Many studies have shown that more educated labor forces prefer a greener environment. Kahn and Walsh (2015) also found that as the mobility of labor and enterprises increases, the quality of the environment has an increasingly important impact on residents' choices of cities to live in. A high-quality living environment attracts labor inflow, correspondingly, a polluted environment causes labor outflow. Chen et al. (2017) found that air pollution significantly increases local population outflow, and the resulting high-skilled labor outflow has a significant impact on the economy of the entire region.

From the literature in this section, it can be seen that China's urbanization is closely related to industrialization. The process of industrialization provides new job opportunities for cities, thereby attracting the rural population to migrate to urban areas, and then expanding urban development. The improvement of urban development will help industrial development through the agglomeration effect. Therefore, in the early stages of urban development, job opportunities provided by industrialization are the main reason for residents' choice of work location. With the deepening understanding of environmental protection among residents and the increase in literature on the health hazards of environmental pollution, for more developed cities and more educated labor forces, the ecological environment factor is becoming more and more important in residents' choice of residence, which also provides support for the study of the relationship between environmental regulation and urbanization in the following text.

## 3. Theoretical Analysis

This chapter aims to study the relationship between environmental regulation and urbanization rate. Typically, the process of urbanization rate increasing involves the migration of residents from rural areas to urban areas. However, in China, urban population mobility is not entirely free, as the household registration system (hukou) restricts the migration of rural labor to cities (Peng & Guo, 2007). The hukou system is a crucial institution in the development of Chinese cities. The urbanization rate is generally defined as the proportion of non-agricultural hukou holders in the total population. Therefore, even if the number of people living in cities increases, the urbanization rate may not rise due to the difficulty new residents face in obtaining hukou. In 2005, China's urbanization rate, calculated based on non-agricultural hukou holders, was about 28%, while the proportion of the permanent urban population reached 43%, with migrant workers accounting for a significant part of this difference (Sheng, 2007).

Additionally, whether a person has a hukou in a city significantly affects their quality of life. The hukou system creates barriers for migrant workers in terms of geography, employment, access to public services, and administrative power. Migrant workers tend to work longer hours than local workers but earn only 61% of their wages (Meng & Zhang, 2006). Although farmers can now move freely without hukou restrictions, local governments implement employment policies to protect local hukou holders (Cai et al., 2001). Tian (2010)found that the hukou threshold is the main obstacle for migrant workers to obtain high-income jobs within the public system, and Wu and Zhang (2014) found that the hukou system causes occupational segregation between migrant workers and local urban workers.

In terms of public services, most affordable housing application requirements in cities include local hukou. Non-local hukou residents also have lower social welfare standards compared to local hukou holders. Furthermore, hukou affects the education of residents' children, especially at the high school level, where many non-local hukou children must return to their hukou location to take the college entrance exam. Although the State Council has issued policies requiring education and employment opportunities not to be linked to hukou, the actual effects are not significant. The above descriptions indicate that obtaining local hukou greatly affects residents' quality of life and ultimately influences their choice of work location. Additionally, the hukou policy in China has gradually diversified across provinces. In first-tier cities like Beijing and Shanghai, hukou thresholds are increasingly high, while second and third-tier cities have increasingly relaxed hukou thresholds. Wang et al. (2010) proposed through an optimal decision model that, ideally, hukou thresholds in small and medium-sized cities will gradually disappear, while in first-tier cities, the thresholds will stabilize at a high level. The combination of differentiated hukou policies and environmental regulations will have varied and even opposite effects on urban population mobility in different regions of China.

3.1. Crowding-Out Effect

Previous literature has shown that environmental regulation affects the migration and employment of enterprises, with job opportunities becoming the most critical factor in population mobility (Ding et al., 2005). This section will analyze whether environmental regulation could cause urban population outflow by shutting down pollution-intensive enterprises or prompting their relocation, thereby creating a crowding-out effect on local urbanization development.

In underdeveloped areas, industry remains an important factor for urban development. In less developed cities with lower hukou thresholds, the proportion of workers in COD-related pollution enterprises is higher in the total urban labor force, and most workers can obtain local urban hukou. Figure 1 depicts the relationship between the proportion of employment in pollution industries and per capita GDP at the prefecture-level from 2003-2008. The fitted dashed line shows a significant downward trend, indicating that the proportion of employment in pollution enterprises increases as per capita GDP decreases. If the "Eleventh Five-Year Plan" leads to stricter environmental regulations, resulting in the closure or relocation of pollution factories, it will inevitably prompt some workers to migrate or seek new job opportunities. Since local hukou significantly impacts living standards, workers are likely to move along with the factories, taking their hukou with them, which negatively impacts the urbanization rate.

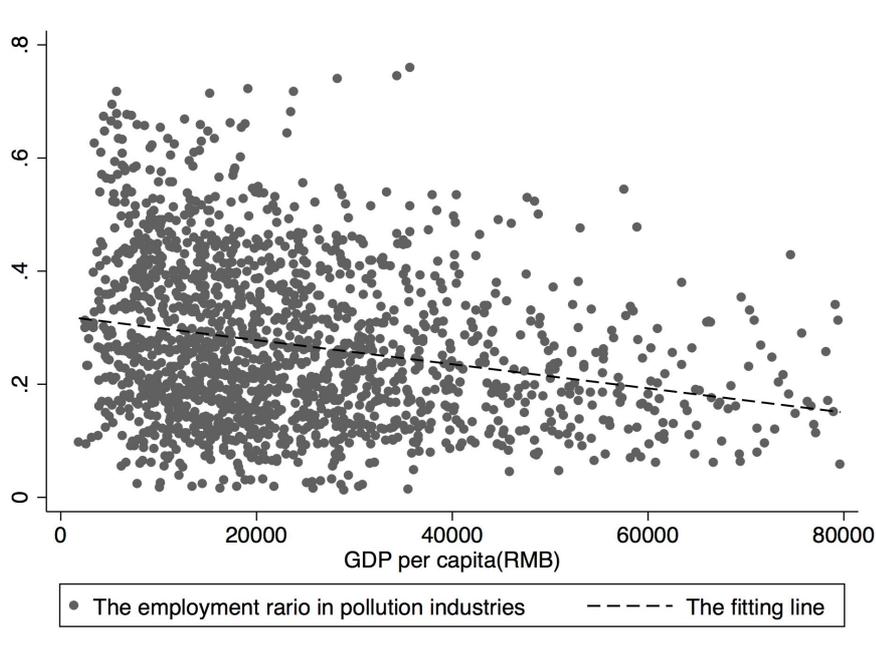

Figure 1 The relationship between Per capita GDP and the proportion of employment in polluting industries

Conversely, for cities with relatively lax regulations, which are generally underdeveloped areas, the influx of new factories will create new job opportunities, and incoming labor will likely obtain local hukou, promoting urbanization. Thus, environmental regulation can have a crowding-out effect on urban population mobility, particularly in underdeveloped areas, suppressing the increase in urbanization rate.

In contrast, in more developed regions like the east, finance and real estate, those high-end services increasingly dominate population inflow (Yu, 2004). In these developed cities with high hukou thresholds, the proportion of COD-related pollution enterprises is already low. Workers in these factories are usually less educated and find it difficult to obtain local hukou. The fifth national census data shows that at least half of the second industry employees are migrant workers (China Agricultural Issue Drafting Group, 2006). Therefore, even if environmental regulations cause enterprises and labor to migrate, it does not significantly impact the urbanization rate of developed cities. Overall, this section posits that environmental regulation will significantly crowd out urban population mobility in underdeveloped western cities, while the effect may be negligible in developed eastern cities, leading to Hypothesis 1:

Hypothesis 1: In underdeveloped cities, environmental regulation will have a significant negative impact on their urbanization rate.

3.2. Crowding-In Effect

The Yangtze River Delta, Pearl River Delta, and Beijing-Tianjin-Hebei regions have attracted populations from across the country over the past 20 years, becoming major urban agglomerations in China. For developed eastern cities, the impact of environmental regulation on population mobility differs significantly from that in the west. First, the hukou threshold is higher in developed cities, and highly educated people typically do not work in pollution-intensive industries. Moreover, in these cities, workers in pollution enterprises usually do not have local hukou. Therefore, environmental regulation does not significantly crowd out urbanization. As cities develop, highly educated labor increasingly considers urban environmental quality when choosing work locations. Yu (2004) noted that living quality has gradually become a driving factor for population inflow in the east, with new residents mainly engaged in high-end services or knowledge-based industries. Glaeser, Kolko and Saiz (2000) found that as enterprise mobility increases, urban development increasingly depends on the benefits it offers residents. He found that cities with better infrastructure and environments develop faster, and the cost of living in good cities rises more than wages, providing additional benefits to residents.

Figure 2 illustrates the relationship between the proportion of employment in non-pollution industries and per capita GDP at the prefecture-level from 2003-2008. The fitted dashed line slopes upwards, indicating that the proportion of employment in non-pollution enterprises increases as per capita GDP rises, consistent with previous analysis.

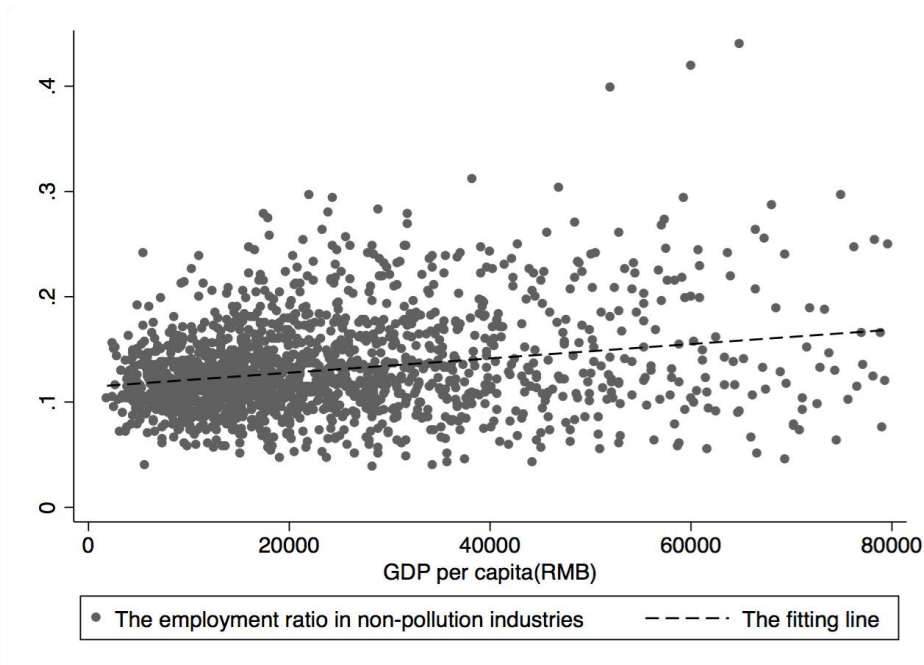

Figure 2 The relationship between Per capita GDP and the proportion of employment in non-polluting industries

Based on the above analysis, this section anticipates that for developed cities, environmental regulation will attract highly educated labor by improving local living environments. This labor typically meets local hukou standards, increasing the urbanization rate. Moreover, since most high-end labor is engaged in high-end services, environmental regulation is expected to significantly positively impact the proportion of service industry workers. Thus, Hypothesis 2 is proposed:

Hypothesis 2: In developed cities, environmental regulation will have a significant positive impact on their urbanization rate.

# 4. Econometric Model and Data Sources

## 4.1. Econometric Model

The primary objective of the empirical study in this chapter is to identify the impact of the "Eleventh Five-Year Plan" COD emission restriction policy on urban population mobility. Based on the analysis above, this study posits that environmental regulation has two effects on urban population mobility: crowding-in and crowding-out effects. In more developed regions, environmental regulation improves urban environments and employment in non-polluting industries, promoting urban development through the crowding-in effect. In less developed regions, environmental regulation reduces production and factory employment or relocates factories, inhibiting urban development through the crowding-out effect. The varying implementation strengths of the COD emission restriction policy across provinces during the "Eleventh Five-Year Plan" provide the conditions for this study's difference-in-differences model.

This chapter focuses on the environmental regulation of the "Eleventh Five-Year Plan" COD emission restriction policy. Similar to energy consumption regulation, the central government set COD reduction targets for different provinces, and then provincial governments further allocated these targets to their respective prefecture-level cities. The variation in provincial COD reduction targets, with more stringent restrictions for provinces with more severe pollution, forms the basis for the difference-in-differences model used in this study. Thus, in this model, the control group consists of prefecture-level cities with relatively light regulation, while the treatment group consists of those with stricter regulation. Based on the theoretical analysis, environmental regulation has two opposing effects on urban population mobility: crowding-in and crowding-out. If the overall crowding-in effect is greater, it indicates that environmental regulation helps increase the urbanization rate, and vice versa. If the coefficient of the difference-in-differences model is not significant, it suggests that the crowding-in and crowding-out effects may offset each other overall.

Based on the discussion of the relationship between environmental regulation and urban population mobility, the difference-in-differences model in this chapter is as follows:

$$REG_{i,t} = \beta Reg_p \times Post_t + X'_{i,t}\theta + \gamma_t + \eta_i + \varepsilon_{i,t} \quad (1)$$

where *i* and *t* represent prefecture-level cities and years, respectively. *URB* represents the urbanization rate of the municipal district. *X* includes control variables at the prefecture-level that affect the urbanization rate, such as population and actual GDP. *Reg* measures the strictness of COD regulation, approximated by the COD reduction targets for prefecture-level cities derived from provincial targets (see table 3). *Post* is a dummy variable indicating the regulation treatment period, equal to 1 if *t*>2006 and 0 otherwise. $\eta_i$ represents city fixed effects, capturing all unobservable permanent effects at the city level; $\gamma_t$ represents year fixed effects, capturing unobservable time-varying factors such as monetary and fiscal policies, business cycles, and macroeconomic shocks; and $\varepsilon_{i,t}$ is the error term. The coefficient of interest is *β*, representing the overall impact of environmental regulation on the national urbanization rate.

This study uses data from prefecture-level cities to decompose the effects of environmental regulation on urban population mobility into crowding-in and crowding-out effects. Since these effects might offset each other, making the overall sample regression coefficient insignificant, this study seeks to separately identify the crowding-out and crowding-in effects. The crowding-out effect primarily targets less developed cities and regions with lower per capita GDP, where pollution industries employ a higher proportion of the workforce, and local hukou holders are more common. The crowding-in effect primarily targets more developed eastern regions with higher per capita GDP, where high hukou thresholds attract highly educated labor. Thus, this study uses regional classifications of east, central, and west, as well as income-based classifications, for heterogeneous regression to attempt to distinguish these effects. Generally, the eastern, central, and western regions of China differ significantly in urban development, per capita GDP, and geographical features. The eastern and central regions are relatively developed, while the western region is less developed. Additionally, geographical environments and cultures within each region are quite similar, making this classification method widely used by scholars (Poumanyvong & Kaneko, 2010). Therefore, this chapter conducts regressions for the eastern, central, and western regions to try to separate the crowding-in and crowding-out effects.

Furthermore, due to development imbalances within eastern, central, and western regions, these imbalances might affect our expected decomposed results. Therefore, this study also classifies all prefecture-level cities into three categories based on per capita GDP: below 21,044 yuan, between 21,044 and 31,906 yuan, and above 31,906 yuan, and performs separate regressions for each category. The results obtained from the above analysis are the sum of the crowding-in and crowding-out effects, without separately analyzing each effect. To examine the crowding-in and crowding-out effects from a more micro perspective, this study also changes the dependent variable from the urbanization rate to polluting and non-polluting industries. By doing so, this study analyzes urban population mobility from the perspective of industries. To reduce biases from omitted variables, this study also controls for individual fixed effects, time fixed effects, and other variables that might influence the urbanization rate. The descriptive statistics of related variables are shown in table 1 and talbe 2. The urbanization rate here represents the urbanization rate of prefecture-level cities.

Table 1 Variable Definitions

| Variables | Definition | Unit |
|---|---|---|
| URB | Urban's share of non agriculture population in total population | % |
| P | Population | 10 thousand |
| A | GDP per capita of the city | RMB |
| REG | COD reduction target of prefecture-level cities | 10 thousand tons |
| Post2006 | Dummy variable | |
| Pol-ratio | The proportion of employment in COD-related pollution industries | % |
| Nonpol-ratio | The proportion of employment in Non COD-related pollution industries | % |

Table 2 Summary Statistics

| Variables | Obs. | Mean | Std. Dev. | Min | Max |
|---|---|---|---|---|---|
| URB | 1,644 | 0.602 | 0.246 | 0.122 | 1 |
| P | 1,683 | 127.1 | 156.9 | 14.08 | 1534 |
| A | 1,683 | 24798 | 17157 | 1847 | 100216 |
| REG | 1,683 | 0.421 | 0.66 | 0 | 4.372 |
| Post2006 | 1,683 | 0.329 | 0.47 | 0 | 1 |
| Pol-ratio | 1,683 | 0.268 | 0.141 | 0.0113 | 0.759 |
| Nonpol-ratio | 1,683 | 0.131 | 0.0456 | 0.384 | 0.439 |

4.2. Data Sources

The urbanization rate data used in this study is from the "China Population and Employment Statistical Yearbook" from 2003 to 2008. Control variables for prefecture-level cities mainly come from each city's statistical yearbooks. The environmental regulation variable is derived from the "Total Emission Control Plan for Major Pollutants During the Eleventh Five-Year Plan".

To quantitatively study the impact of environmental regulation on the urbanization rate, consistent with Chen et al. (2018), this study decomposes environmental regulation to the prefecture-level. In the "Eleventh Five-Year Plan", the central government's water pollution control targets only include COD emissions. COD is one of the most severe pollutants in water pollution. The central government first set a national overall COD reduction target, requiring a 10% reduction, equivalent to 1.41 million tons, by the end of 2010. Subsequently, in August, the central government distributed the national targets to each province. Each province's COD reduction target was primarily determined by its economic development, industrial structure, and current environmental pollution levels. Similar to the "Eleventh Five-Year Energy Policy", provinces with more severe pollution received stricter regulations. Table 3 represents the COD reduction targets for each province. These targets are binding indicators, influencing the promotion of local government officials.

Table 3 "Eleventh Five -Year Plan" planned COD emission reduction goals in each province

| Province | 2005 discharge | 2010 discharge target | Reduction ratio(%) | Province | 2005 discharge | 2010 discharge target | Reduction ratio(%) |
|---|---|---|---|---|---|---|---|
| Total | 1414.2 | 1263.9 | -10.6 | Shandong | 77 | 65.5 | -14.9 |
| Beijing | 11.6 | 9.9 | -14.7 | Henan | 72.1 | 64.3 | -10.8 |
| Tianjing | 14.6 | 13.2 | -9.6 | Hubei | 61.6 | 58.5 | -5 |
| Hebei | 66.1 | 56.1 | -15.1 | Hunan | 89.5 | 80.5 | -10.1 |
| Shanxi | 38.7 | 33.6 | -13.2 | Guangdong | 105.8 | 89.9 | -15 |
| Inner Mongolia | 29.7 | 27.7 | -6.7 | Shenzhen | 5.59 | 4.47 | -20 |
| Liaoning | 64.4 | 56.1 | -12.9 | Guangxi | 107 | 94 | -12.1 |
| Dalian | 6.01 | 5.05 | -16 | Hainan | 9.5 | 9.5 | 0 |
| Jilin | 40.7 | 36.5 | -10.3 | Chongqing | 26.9 | 23.9 | -11.2 |
| Heilongjiang | 50.4 | 45.2 | -10.3 | Sichuan | 78.3 | 74.4 | -5 |
| Shanghai | 30.4 | 25.9 | -14.8 | Guizhou | 22.6 | 21 | -7.1 |
| Jiangsu | 96.6 | 82 | -15.1 | Yunnan | 28.5 | 27.1 | -4.9 |
| Zhejiang | 59.5 | 50.5 | -15.1 | Tibet | 1.4 | 1.4 | 0 |
| Ningbo | 5.22 | 4.44 | -14.9 | Shaanxi | 35 | 31.5 | -10 |
| Anhui | 44.4 | 41.5 | -6.5 | Gansu | 18.2 | 16.8 | -7.7 |
| Fujian | 39.4 | 37.5 | -4.8 | Qinghai | 7.2 | 7.2 | 0 |
| Xiamen | 5.56 | 4.94 | -11.2 | Ningxia | 14.3 | 12.2 | -14.7 |
| Jiangxi | 45.7 | 43.4 | -5 | Xinjiang | 27.1 | 27.1 | 0 |

Note: 1. The total control target of 10% of the national chemical oxygen demand is 12.728 million tons, which is actually distributed to 12.639 million tons in each province, and the nation state reserves 89,000 tons to achieve chemical oxygen -demand sewage discharge rights paid distribution and transaction pilot work. 2. Chemical Oxygen Demand of the Xinjiang Production and Construction Corps does not include the source of daily life from various parts of the Corps and the chemical oxygen demand of the Agricultural Eighth Division (Shihezi City).

In November 2006, the Ministry of Environmental Protection issued a provincial-level COD reduction target allocation guide, specifying the amount of COD each city must reduce. This guide also provided a clear formula for calculating the COD reduction requirements for each prefecture-level city within a province. The formula is as follows:

$$\Delta COD_c = \Delta COD_p \times \frac{P_c}{\sum_{i=1}^{I} P_i} \quad (2)$$

where $\Delta COD_c$ represents the COD reduction required for the prefecture-level city from 2006 to 2010. On the right side of the equation, $\Delta COD_p$ represents the COD reduction required for the province to which the city belongs from 2006 to 2010. $P_c$ represents the COD emissions of the prefecture-level city in 2005. $i$ represents all prefecture-level cities within the province. Equation (2) allocates provincial COD reduction targets based on the proportion of COD emissions from each prefecture-level city.

Ideally, the COD regulation intensity should be calculated according to equation (2). However, since COD emissions from enterprises are not directly regulated but estimated by local governments based on each enterprise's output, this study, following Chen et al. (2018), uses the output proportion of related industries in each prefecture-level city to approximate its COD emission proportion. The specific formula is as follows:

$$\Delta COD_c = \Delta COD_p \times \sum_{i=1}^{39} \mu_i \frac{O_{i,c}}{O_{i,p}} \quad (3)$$

where the second term on the right represents the proportion of water pollution enterprise output in the prefecture-level city relative to the province. We then sum the weighted outputs of 39 two-digit industries, with $\mu_i$ representing the proportion of each industry within the province.

To meet the reduction requirements of the "Eleventh Five-Year Plan", local governments typically have two measures: issuing pollution discharge permits and post-discharge monitoring. If violations are found, the relevant factories will be warned, fined, or have their business licenses revoked. Factories must meet stringent discharge standards and have more advanced pollution treatment equipment to obtain discharge permits. Local governments can also directly close heavily polluting factories or control the entry of new polluting enterprises through land use restrictions. All these measures increase production costs for enterprises. If production costs become too high, enterprises may choose to close factories and relocate to regions with relatively lenient regulations.

China's infrastructure development also facilitates the relocation of factories across regions. (Alder et al., 2016; Chen, Oliva & Zhang, 2017) all indicate that many local governments in China establish industrial parks to attract investment, providing highways, sewage treatment systems, and other public facilities, greatly facilitating factory relocation. Moreover, the average equipment utilization rate of Chinese enterprises is very low, allowing enterprises to change their production behavior by adjusting the investment scale in different factories rather than relocating entire factories to respond to the impact of environmental regulations (Shen & Chen, 2017).

Generally, environmental regulations are multidimensional and difficult to quantify (Shadbegian & Wolverton, 2010). To quantify the intensity of environmental regulations, especially water pollution regulations, many scholars use ex-post quantification methods, such as increased taxes and actual expenditure on pollution reduction (Gray & Shadbegian, 2003). Wu et al. (2016) adopted an ex-ante quantification method for water pollution regulation, analyzing the impact of provincial COD reduction regulations on the location choices of new factories. Ex-ante quantification of regulation intensity is preferable to ex-post quantification because ex-post methods are more susceptible to changes made by enterprises in response to varying regulation intensities. Additionally, considering endogeneity, central government regulations are less likely to be influenced by local governments. Therefore, this study selects the COD environmental regulation targets in the "Eleventh Five-Year Plan" as the research object for regulation.

## 5. The empirical result

### 5.1. Regression Results

This section aims to verify Hypotheses 1 and 2. According to the theoretical analysis in Section 4, environmental regulation has both crowding-in and crowding-out effects on urban population mobility. The crowding-out effect refers to the suppression of production and labor demand in pollution-related industries due to increased production costs caused by environmental regulation, leading to an outflow of urban population. The crowding-in effect, on the other hand, refers to the improvement of urban environments in regions with stricter environmental regulations, attracting population inflows. Additionally, environmental regulation can promote innovation and encourage the development of non-polluting industries, also attracting population inflows. For each city, both crowding-in and crowding-out effects exist simultaneously. The relative magnitude of these effects depends on the city's development stage, industrial structure, and the intensity of environmental regulation it faces. Ultimately, the impact of environmental regulation on the urbanization rate is the result of the combined effects of crowding-in and crowding-out.

This section uses data from Chinese prefecture-level cities from 2003 to 2008 to analyze the crowding-in and crowding-out effects of environmental regulation on the urbanization rate. Table 4 reports the regression results of the difference-in-differences model for the entire country. Model 1 includes only control variables at the prefecture level and does not control for fixed effects. To minimize endogeneity issues caused by omitted variables, Model 2 adds time fixed effects to account for macroeconomic factors changing over time. Model 3 further adds individual fixed effects to Model 2 to absorb unobservable factors that remain constant at the prefecture level over time. In all models, the coefficient of Regulation*Post is not significant. According to the theoretical analysis, this suggests that the crowding-in and crowding-out effects of environmental regulation on urban population mobility may offset each other nationwide. To obtain more meaningful results and verify the crowding-in and crowding-out effects discussed in the theoretical analysis, this section conducts heterogeneity analysis by dividing the basic model into two different types.

First, this section divides the country into eastern, central, and western regions. Table 5 reports the regression results of the difference-in-differences model for these three regions separately. After controlling for prefecture-level control variables and individual and time fixed effects, the coefficients for the eastern and central regions are positive but not statistically significant. Only the western region has a significantly negative coefficient, indicating that environmental regulation has a greater crowding-out effect on urban population mobility in the western region, ultimately leading to a decrease in the urbanization rate. Compared to the western region, the eastern region has better urban development and higher per capita income. Based on the theoretical analysis, the crowding-in effect of environmental regulation on urban population mobility should be greater than the crowding-out effect in the eastern and central regions, while the crowding-out effect should be greater in the western region. However, the coefficient for the eastern region is not significant. Specifically, there are also underdeveloped prefecture-level cities with low per capita income in the eastern and central regions. Figure 3 shows the distribution of per capita GDP in the eastern, central, and western regions. Although the eastern region's average per capita GDP is significantly higher than that of the central and western regions, there are also some cities with high per capita GDP and developed urban areas in the central and western regions, and some underdeveloped cities in the eastern region. This may explain the insignificant coefficient and suggests that dividing the country solely into eastern, central, and western regions may not be reasonable. Therefore, to further verify our hypothesis, this chapter classifies prefecture-level cities into three categories based on per capita income: high, medium, and low-income cities, and then performs difference-in-differences regressions for each category.

Table 4 The effect of regulation on urbanization

| Variables | Model(1) | Model(2) | Model(3) |
| --- | --- | --- | --- |
|  | URB | URB | URB |
| Regulation*Post | -0.0115 | 0.0204 | 0.0025 |
|  | (0.024) | (0.0253) | (0.0061) |
| City controlled variables | Y | Y | Y |
| Year fixed effect | N | Y | Y |
| City fixed effect | N | N | Y |
| Obs. | 1,663 | 1,663 | 1,663 |
| $R^2$ | 0.208 | 0.224 | 0.979 |

Notes: *** stands for p < 0.01, ** stands for p < 0.05, and * stands for p < 0.1. Robust standard errors are reported in parentheses.

Table 5 Heterogeneity test 1: The effect of regulation on urbanization

| Variables | Model(east) | Model(middle) | Model(west) |
|---|---|---|---|
|  | URB | URB | URB |
| Regulation*Post | 0.00889 | 0.000341 | -0.0338*** |
|  | (0.00949) | (0.016) | (0.00965) |
| City controlled variables | Y | Y | Y |
| Year fixed effect | Y | Y | Y |
| City fixed effect | Y | Y | Y |
| Obs. | 565 | 600 | 498 |
| $R^2$ | 0.956 | 0.975 | 0.99 |

Notes: *** stands for p < 0.01, ** stands for p < 0.05, and * stands for p < 0.1. Robust standard errors are reported in parentheses.

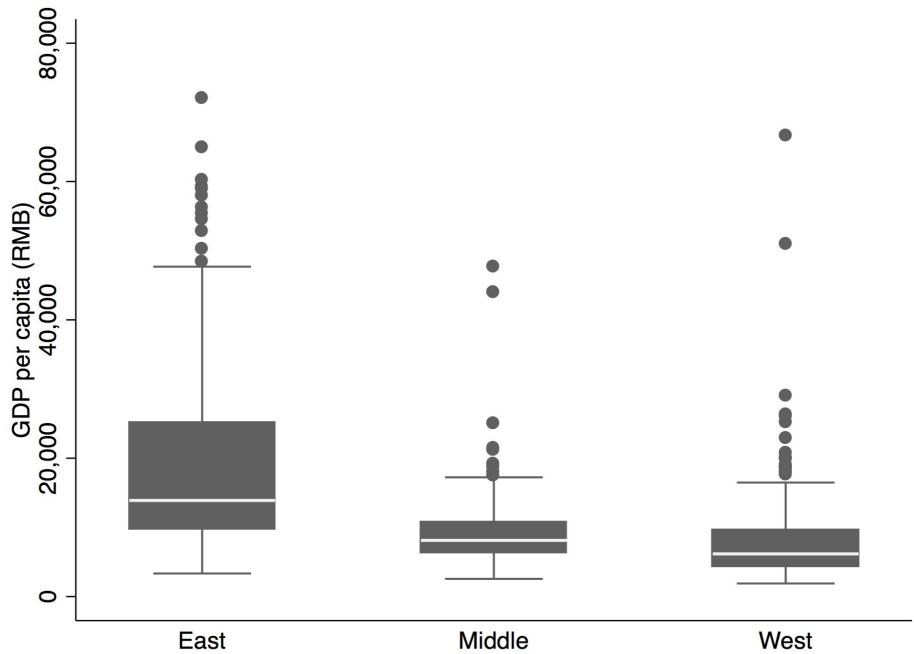

Figure 3 Per capita GDP distribution in different regions

Using the World Bank's standards, Chinese cities are classified by per capita GDP as follows: below 25,000 yuan, between 25,000 and 70,000 yuan, and above 70,000 yuan, corresponding to low, medium, and high-income cities, respectively. Since most Chinese cities fall into the low or medium-income categories, strictly following the World Bank's classification would result in very few high-income city samples. Therefore, based on China's actual development situation, this chapter uses the 50th (21,044 yuan) and 75th (31,906 yuan) percentiles of per capita GDP as thresholds for classifying cities into low, medium, and high-income categories. Heterogeneity regressions are conducted based on both classification standards. This section expects that for high-income cities, environmental regulation has a significant positive impact on urbanization rates through crowding-in effect. The regression results are shown in Tables 6 and 7.

In Table 6, Models 1 to 3 represent the regression results based on the classification of Chinese cities by per capita GDP percentiles. Table 7 reports the results based on the World Bank's income classification standards. For both classification standards, the coefficient for low-income cities is negative and statistically significant at the 1% level, indicating that environmental regulation has a significant crowding-out effect on low-income cities, suppressing the urbanization rate. Additionally, the absolute value of the coefficient in Model 1 of Table 7 is greater than that in Model 1 of Table 6, suggesting that environmental regulation also has a significant negative impact on the urbanization rate of cities with per capita GDP between 21,044 and 25,000 yuan. For medium-income cities, the coefficients of environmental regulation are not significant, indicating that crowding-in and crowding-out effects offset each other during the mid-stage of urban development, having no significant impact on the urbanization rate. Comparing the results under the two standards, the coefficients shift from negative to positive, indicating that as per capita GDP rises, the impact of environmental regulation on the urbanization rate becomes more positive. Finally, for high-income cities, the coefficients of environmental regulation are positive and statistically significant, indicating that environmental regulation mainly exhibits a crowding-in effect on high-income cities, ultimately promoting the urbanization rate. Based on these tables, it can be inferred that environmental regulation has a more significant crowding-in effect on cities with per capita GDP above approximately 30,000 yuan. These results are consistent with the theoretical analysis in this section. Next, robustness checks are performed on the significant regression results.

Table 6 Heterogeneity test 2:The effect of regulation on urbanization

| Variables | Model(1) (A<21044) URB | Model(2) (31906>A>21044) URB | Model(3) (A>31906) URB |
|---|---|---|---|
| Regulation*Post | -0.0439*** | -0.0209 | 0.0159** |
|  | (0.0169) | (0.0247) | (0.00748) |
| City controlled variables | Y | Y | Y |
| Year fixed effect | Y | Y | Y |
| City fixed effect | Y | Y | Y |
| Obs. | 890 | 375 | 398 |
| R² | 0.987 | 0.958 | 0.968 |

Notes: *** stands for p < 0.01, ** stands for p < 0.05, and * stands for p < 0.1. Robust standard errors are reported in parentheses.

Table 7 Heterogeneity test 3:The effect of regulation on urbanization

| Variables | Model(1) (A<25000) URB | Model(2) (70000>A<25000) URB | Model(3) (A>70000) URB |
|---|---|---|---|
| Regulation*Post | -0.0649*** | 0.01 | 0.103*** |
|  | (0.0145) | (0.0085) | (0.0273) |
| City controlled variables | Y | Y | Y |
| Year fixed effect | Y | Y | Y |
| City fixed effect | Y | Y | Y |
| Obs. | 1,046 | 567 | 50 |
| R² | 0.983 | 0.968 | 0.983 |

Notes: *** stands for p < 0.01, ** stands for p < 0.05, and * stands for p < 0.1. Robust standard errors are reported in parentheses.

## 5.2. Parallel Trend Assumption Test

An important issue with the difference-in-differences model is that the different characteristics exhibited by cities after policy implementation may be due to pre-existing factors before the policy was implemented. To ensure the robustness of the results in the difference-in-differences model, it is important to verify the parallel trend assumption. This assumption requires that the trends in urbanization between the control and treatment groups should be parallel before the implementation of the "Eleventh Five-Year Plan" environmental policy, controlling for city-level variables. The "Eleventh Five-Year Plan" environmental regulation was implemented in 2006, so the parallel trend test is given by the following equation:

$$URB_{i,t} = \sum_{k=-3}^{2} Reg_p * Year_{2006+k} \beta_k + X'_{i,t}\theta + \gamma_t + \eta_i + \varepsilon_{i,t} \quad (4)$$

where $URB_{i,t}$ represents the urbanization rate of city $i$ in year $t$. $Reg_p$ represents the stringency of COD reduction regulation in the "Eleventh Five-Year Plan". $Year_{2006+k}$ represents year dummy variables. Note that 2003 is omitted in equation (4) as it is the reference year for the regression model. The regression results of the parallel trend test are reported in Table 8. To better visualize the analysis, the estimated coefficients and their 90% confidence intervals from the parallel trend regression are plotted in Figures 4 to 6. Both the figures and the table show that the coefficients are not significant in the two years before the implementation of environmental regulation. After the regulation came into effect, the coefficients become significant and negative. These results are consistent with the parallel trend assumption in the difference-in-differences analysis, supporting the empirical analysis based on this method.

Table 8 Parallel trend assumption test

| Variables | Model(west) | Model(Cities of low GDP per capita) | Model(Cities of high GDP per capita) |
|---|---|---|---|
| | URB | URB | URB |
| Regulation* Year2004 | 0.00802 | -0.00742 | 0.00982 |
| | (0.0164) | (0.0152) | (0.0121) |
| Regulation* Year2005 | -0.0269 | -0.0172 | 0.00541 |
| | (0.0167) | (0.0166) | (0.0120) |
| Regulation* Year2006 | -0.0429*** | -0.0194 | 0.0101 |
| | (0.0164) | (0.0170) | (0.0118) |
| Regulation* Year2007 | -0.0385** | -0.0601*** | 0.0266** |
| | (0.0164) | (0.0198) | (0.0116) |
| Regulation* Year2008 | -0.0371** | -0.0599** | 0.0285** |
| | (0.0170) | (0.0242) | (0.0118) |
| City controlled variables | Y | Y | Y |
| Year fixed effect | Y | Y | Y |
| City fixed effect | Y | Y | Y |
| Obs. | 498 | 1,265 | 232 |
| $R^2$ | 0.99 | 0.981 | 0.984 |

Notes: *** stands for $p < 0.01$, ** stands for $p < 0.05$, and * stands for $p < 0.1$. Robust standard errors are reported in parentheses.

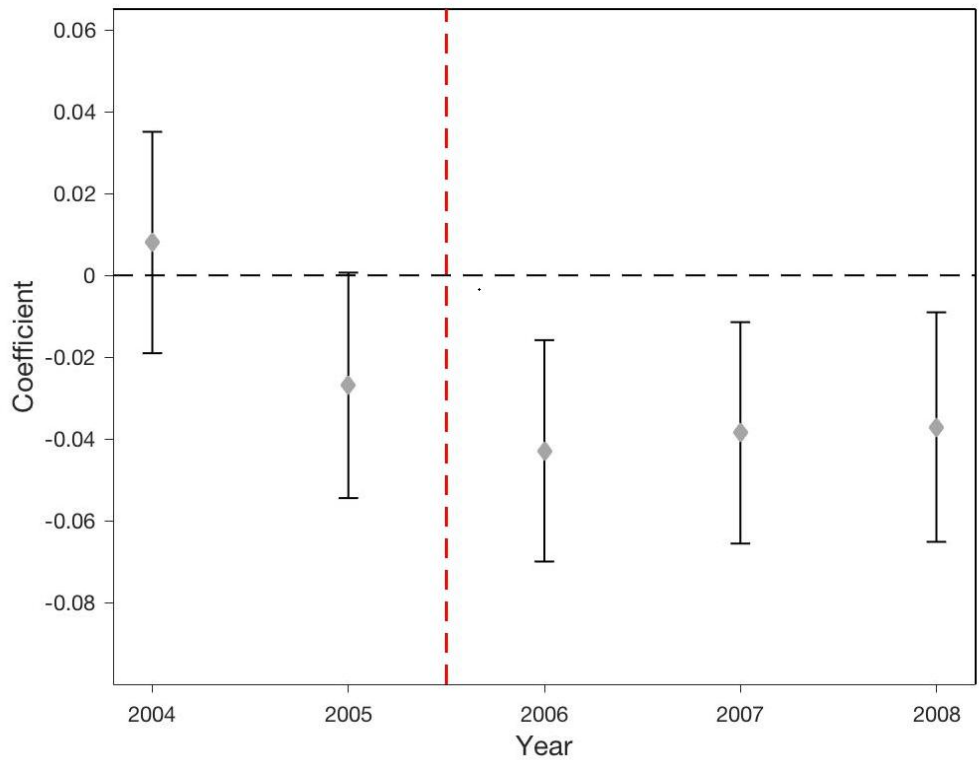

Figure 4 West：Parallel Trend Assumption Test

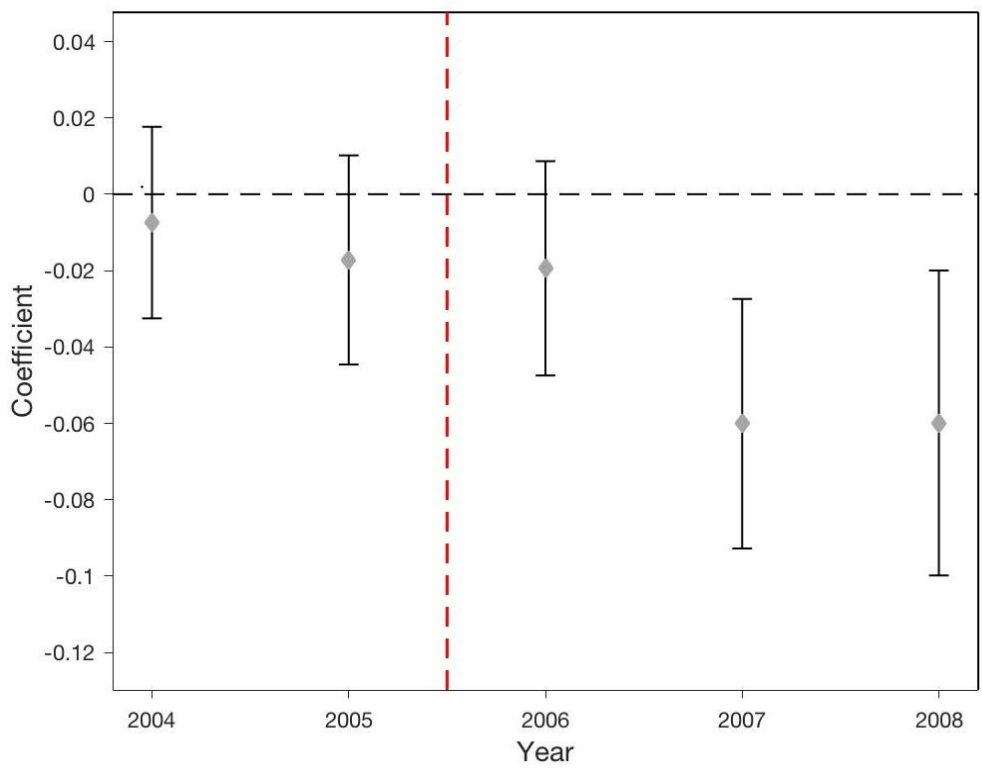

Figure 5 Low-income region：Parallel Trend Assumption Test

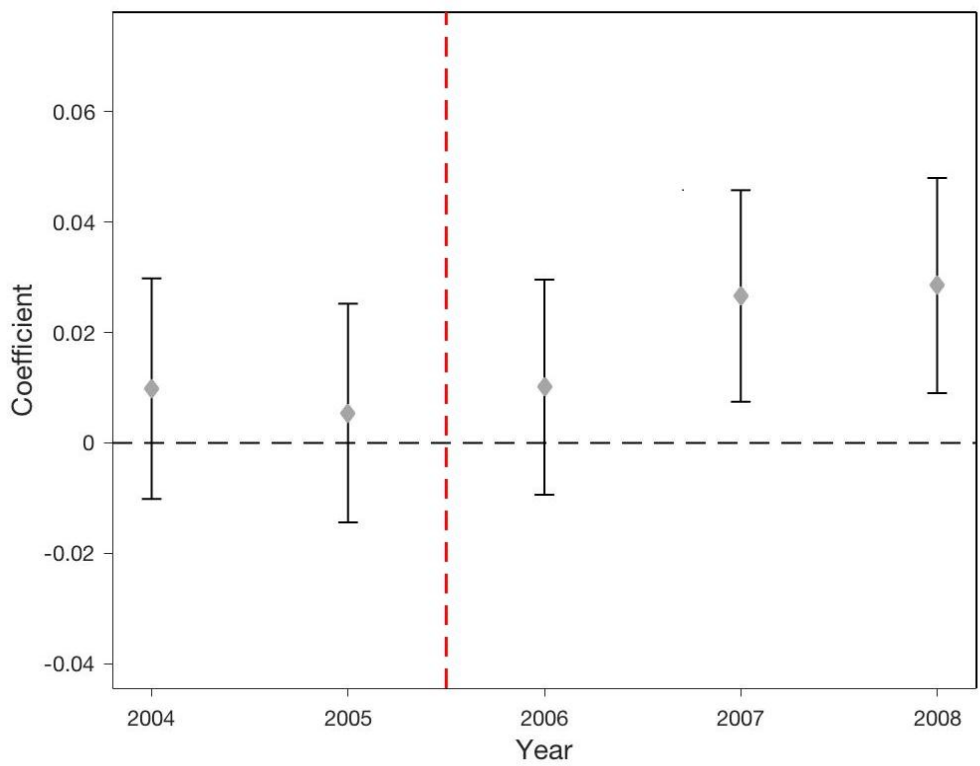

Figure 6 High-income region：Parallel Trend Assumption Test

## 5.3. Mechanism Discussion: The Impact of Environmental Regulation on the Employment Distribution Across Industries

The basic regression results from the previous section indicate that environmental regulation has a significant positive impact on the urbanization rate in high-income cities, a significant negative impact in low-income cities, and no significant impact in medium-income cities. Changes in the urbanization rate of a city are certainly related to population migration, which is largely driven by employment choices. According to the theoretical analysis in Section 3, environmental regulation has a crowding-out effect on the urbanization rate of pollution-intensive cities and a crowding-in effect on high-income cities.

This section first examines the proportion of employment in pollution industries versus non-pollution industries across prefecture-level cities. The seven two-digit industries with the most severe COD pollution are paper products, textiles, agricultural and sideline food processing, chemical raw materials and chemical products manufacturing, beverage manufacturing, food manufacturing, and pharmaceutical manufacturing. This section defines non-polluting industries within the service sector, such as finance, real estate, scientific research, and computer information, as non-polluting industries. Heterogeneous regression analyses are conducted nationwide for both types of industries. If the theoretical analysis in Section 3 is correct, then environmental regulation impacts urban population mobility by influencing employment in pollution versus non-pollution industries. The regression results are shown in Table 9.

Table 9 Mechanism test

| Variables | Model(1) (COD-related industries) Pol-ratio | Model(2) (COD-related industries) Pol-ratio | Model(3) (Non COD-related industries) Pol-ratio | Model(4) (Non COD-related industries) Pol-ratio |
|---|---|---|---|---|
| Regulation*Post | -0.00514** | -0.00476** | 0.00199*** | 0.0179*** |
| | (0.00254) | (0.00234) | (0.000979) | (0.000993) |
| City controlled variables | N | Y | N | Y |
| Year fixed | Y | Y | Y | Y |

|  | effect | | | |
|---|---|---|---|---|
| City fixed effect | Y | Y | Y | Y |
| Obs. | 1,710 | 1,699 | 1,731 | 1,717 |
| $R^2$ | 0.955 | 0.963 | 0.988 | 0.988 |

Notes: *** stands for $p < 0.01$, ** stands for $p < 0.05$, and * stands for $p < 0.1$. Robust standard errors are reported in parentheses.

In Table 9, Models 1 and 3 represent results without controlling for prefecture-level control variables, while Models 2 and 4 include these controls, such as per capita GDP, population density, and total population. To minimize endogeneity issues caused by omitted variables, consistent with the regression analysis in Section 4, this section controls for individual fixed effects to absorb unobservable factors that remain constant at the prefecture level over time. It also controls for time fixed effects to absorb macroeconomic factors changing over time. The regression results show that the coefficients are significant regardless of whether prefecture-level variables are controlled for, indicating that the regression results are robust. Environmental regulation significantly reduces the proportion of employment in pollution industries (-0.00476) and significantly increases the proportion of employment in non-pollution industries (0.0179). This indirectly confirms that the environmental regulation in the "Eleventh Five-Year Plan" was effectively implemented by local governments.

In summary, Hypotheses 1 and 2 of this chapter are validated.

5.4. Discussion

This paper above discuss the relationship between environmental regulation and urbanization in China. There is also some international cases to support the conclusion of this paper. Bayer et al (2009) study the relationship of migration and air quality in US from 1990-2000 and found that the median household would pay $149-$185 for a one-unit reduction in PM10, which is about as three times as estimated by the conventional model. Walker(2011) study the labor reallocation brought by the Clean Air Act in the United States and found that firms responded to regulation by destroying jobs and has important distributional implication for the affected workforce. The two cases above is consistent with our findings that good environment would attract high-income workforce and the more strict regulation may do harm to the urbanization rate in western area.

Due to the availability of data, our regression is from 2003 to 2008. To justify our regression results from a wider perspective, we find that several cases from other literature could support our conclusion. Many scholars studied the relationship between urbanization and emissions, and find that there is an inverted U-shaped relationship between urbanization and $CO_2$ emissions, which means that $CO_2$ emissions initially rise with urbanization, but beyond the inflection point, emissions fall with urbanization (Martínez-Zarzoso and Maruotti, 2011; Xu et al., 2015; He et al., 2017). Our study suggest that environmental regulation could decrease urbanization through crowding-out effect in the western region and enhance urbanization through crowding-in effect in the high-income areas, which on the hand prove that there is an inverted U-shaped relationship between urbanization and emissions. In the western areas, where a lot of people work in energy-intense factories, environmental regulation could make those factories shut down and move to another less regulated city, which lead those people migrate to another city to find a job and cause the urbanization rate decrease in the former city. In the high-income regions, where the major people work in the service or environmental friendly field, a better living environment could attract people move in and thus increase urbanization rate.

## 6. Political implication

1.Binding target. This paper focuses on environmental regulation, urbanization and population mobility, with the aim of providing policy recommendations and theoretical basis for China's future sustainable, high-quality urban development path. Environmental issues have attracted worldwide attention. In order to protect the environment and curb climate warming, countries around the world have formulated various energy and environmental regulations, but not all regulations can achieve ideal results. Whether the regulation is effective depends first on whether the officials who implement the regulation have sufficient internal motivation to achieve the regulatory goals. Before the "Eleventh Five-Year Plan", the "Tenth Five-Year Plan" also had some requirements for COD emissions, but it was ineffective. The biggest difference is that the "Eleventh Five-Year Plan" listed COD emissions as a binding target, linking the promotion of officials with whether the COD emission target can be achieved. Therefore, in the future formulation of regulations, solving the agency problem and strengthening the internal motivation of local officials to implement regulations are key factors for the success of regulations. In addition, whether the regulatory results can be effectively quantified and supervised is also an important factor affecting the effectiveness of regulations. During the "Eleventh Five-Year Plan", the completion of the COD emissions was supervised every year, which is one of the reasons why the regulation of the "Eleventh Five-Year Plan" was ultimately effective. Therefore, in future regulatory formulation, it is also necessary to consider the quantifiability of the binding goals and strengthen supervision of local governments' implementation of regulations.

2.Collaborative governance. Environmental regulation will affect the location of enterprises, population mobility, and thus affect local economic development. On the other hand, population mobility will also affect local employment, taxation, housing prices and other comprehensive economic factors. This is a systematic problem, not a single issue. The ecological environment is essentially an open evolution, coupled and adaptable complex network system. Ecological environmental governance is a large and complex system engineering, showing the characteristics of gridding, diversification and linkage. Based on this, collaborative governance is considered to be the preferred means to solve ecological and environmental problems. The current focus and difficulty of ecological and environmental governance lies in how to achieve the coupling between the ecological and environmental elements in the ecological and environmental system and the coordination between the ecological and environmental system and the economic and social system. This requires all-round linkage reforms to drive the overall improvement of the ecological and environmental system and its coordination with politics, economy, culture and society. Considering that China's ecological and environmental governance presents a decentralized model between the central government and local governments, in the process of comprehensive green transformation, it is necessary to establish a governance structure and governance mechanism that matches the transformation and development, which requires adjusting the relationship between the central and local governments in the past. On the one hand, it is necessary to increase the assessment of local governments' ecological and environmental protection, encourage local governments to pay attention to ecological and environmental protection, and coordinate the relationship between high-quality development and high-level protection. On the other hand, the central government also needs to promote cooperation in ecological and environmental governance between local governments and within local governments. Therefore, while increasing the assessment of local governments' ecological and environmental protection, the central government adjusts the division of ecological and environmental affairs and expenditure responsibilities between the central and local governments through compensation for local ecological and environmental governance, and realizes incentive compatibility of local ecological and environmental governance.

3. Government and market. Previous text shows that environmental regulation may decrease the urbanization rate and hurt local economy growth in western areas. But this doesn't mean that developing areas should develop economy at the price of pollution. To help western areas to protect the environment and at the same time, develop economy, it could be considered from a long-term perspective and government subsidy perspective.

Ecological and environmental governance helps to improve the local living environment, thereby enhancing the investment value of the region and driving up land prices, alleviating local fiscal pressure. In the process of green transformation, it is necessary to properly handle the relationship between "government" and "market" and promote the combination and connection of effective government and effective market. In the early stage, ecological environmental governance investment is mainly relying on government funding, but ecological environmental governance is a long and difficult process. The government faces huge challenges and pressure in long-term investment. If the value conversion of ecological products cannot be smoothly realized and social capital participation cannot be introduced, ecological environmental governance will be unsustainable. Therefore, in the early stage of overall ecological environmental governance and restoration, the leading and traction role of the government can be played. In the middle and late stages, the focus of governance can be adjusted to incentives, guidance, demonstration and cooperation, retreating from the "first line" to the "second line", from the "front stage" to the "backstage", more support for the participation of market forces and social forces, introducing EOD, POD and ROD models, actively promoting the integrated development of industrial ecology and ecological industrialization, organically combining multiple policy mechanisms, and considering flexible matching and applicability to achieve the best policy results and realize the co-construction, co-governance and sharing of ecological dividends to the greatest extent.

## 7. Conclusion

This paper uses data from prefecture-level cities between 2003 and 2008 to discuss the impact of the "Eleventh Five-Year Plan" environmental regulations on urbanization rates. It first provides a theoretical analysis of the relationship between environmental regulation and urbanization, finding that environmental regulation can influence urban population mobility through both crowding-in and crowding-out effects. Environmental regulation can lead to a crowding-out effect by reducing employment in pollution-related industries, causing urban population outflow. Conversely, it can lead to a crowding-in effect by increasing employment in non-pollution industries, causing urban population inflow. These two effects together influence urban population mobility and ultimately affect urbanization rates.

Empirical results show that, possibly due to the mutual offset of crowding-in and crowding-out effects, the "Eleventh Five-Year Plan" environmental regulations did not have a significant impact on urbanization rates nationwide. To disentangle the crowding-in and crowding-out effects, this paper conducts heterogeneity regressions by region (eastern, central, and western) and by income level (high, medium, and low). The regression results indicate that environmental regulation has a significant negative impact on urbanization rates in the western region and low-income cities, and a significant positive impact on urbanization rates in high-income cities. This suggests that the relative magnitude of crowding-in and crowding-out effects is more closely related to a city's per capita GDP than to its regional classification. Furthermore, by examining the proportion of employment in pollution versus non-pollution industries, it is found that the "Eleventh Five-Year Plan" environmental regulations affect urbanization rates by suppressing employment in pollution industries and promoting employment in non-pollution industries.

Today, energy conservation and environmental protection are global development consensuses, and adhering to urban development strategies is essential for China's progress. Talent is the key driving force behind the development of nations and cities, and highly skilled individuals increasingly consider ecological environments in their choice of residence. In more developed cities, environmental regulation can improve the local ecological environment, attracting highly educated labor and ultimately promoting urban development. Therefore, maintaining strict environmental regulations is an important way to promote sustainable and high-quality development in developed cities.

However, in less developed cities or those relying heavily on traditional industries, environmental regulation may lead to the closure or relocation of local factories, negatively impacting urbanization rates. Previous literature concluded that there is an inverted U-shaped relationship between carbon emissions and urbanization rates. Urban development can help local industries better leverage agglomeration effects, ultimately achieving energy-saving and emission-reduction goals. Although environmental regulation may negatively impact urbanization rates in less developed areas in the short term, it does not mean that strict environmental regulations should not be enforced in these regions. The global emphasis on environmental issues dictates that less developed areas cannot pursue economic growth at the expense of the environment. To better promote urbanization, leverage urban agglomeration advantages, and protect the local ecological environment, it is crucial for less developed regions to develop industries with low or no pollution, utilizing their comparative advantages for urban development.